\documentclass[]{aa}
\usepackage{txfonts}
\usepackage{graphicx}
\newcommand{\Omm}{\Omega_{\rm m}}

\newcommand{\OmL}{\Omega_{\Lambda}}

\newcommand{\OmT}{\Omega_{\rm tot}}

\newcommand{\Ombhh}{\Omega_{\rm b}h^2}

\newcommand{\bc}{\begin{center}}
\newcommand{\ec}{\end{center}}
\newcommand{\bi}{\begin{itemize}}
\newcommand{\ei}{\end{itemize}}
\newcommand{\bt}{\begin{tabular}}
\newcommand{\et}{\end{tabular}}

\newcommand{\be}{\begin{equation}}
\newcommand{\ee}{\end{equation}}
\newcommand{\bd}{\begin{displaymath}}
\newcommand{\ed}{\end{displaymath}}

\newcommand{\bea}{\begin{eqnarray}}
\newcommand{\eea}{\end{eqnarray}}

\newenvironment{rem}[1][]{\begin{center}\begin{tabular}{|p{10cm}|} \hline\begin{small}{\bf #1}}{\end{small}\\ \hline \end{tabular}\end{center}}

\newenvironment{remm}[1][]{\begin{center}\begin{tabular}{||p{10cm}||}\hline\hline\begin{small}{\bf #1}}{\end{small}\\\hline\hline\end{tabular}\end{center}}

\newcommand{\br}{\begin{rem}}
\newcommand{\er}{\end{rem}}
\newcommand{\benc}{\begin{remm}}
\newcommand{\eenc}{\end{remm}}

\newcommand{\lig}{\ \hspace*{5cm} \ \\}

\newenvironment{lignee}[1][]{\begin{center}\begin{tabular}{c} \lig}
{ \hline \\ \end{tabular}\end{center}}
\newcommand{\bl}{\begin{lignee}[]}
\newcommand{\el}{\end{lignee}}

\begin{document}

   \title{Cosmological constraints from Archeops}

   \titlerunning{Cosmological constraints from Archeops}

\author{ 
A.~Beno{\^\i}t\inst{1} \and 
P.~Ade \inst{2} \and 
A.~Amblard \inst{3, \, 24} \and 
R.~Ansari \inst{4} \and 
{{\'E}}.~Aubourg \inst{5, \, 24} \and
S.~Bargot \inst{4} \and
J.~G.~Bartlett \inst{3, \, 24} \and 
J.--Ph.~Bernard \inst{7,\, 16} \and 
R.~S.~Bhatia \inst{8} \and 
A.~Blanchard\inst{6} \and 
J.~J.~Bock \inst{8, \, 9} \and
A.~Boscaleri \inst{10} \and 
F.~R.~Bouchet \inst{11} \and 
A.~Bourrachot \inst{4} \and 
P.~Camus \inst{1} \and 
F.~Couchot \inst{4} \and
P.~de~Bernardis \inst{12} \and 
J.~Delabrouille \inst{3, \, 24} \and
F.--X.~D{\'e}sert \inst{13} \and 
O.~Dor{{\'e} \inst{11}} \and 
M.~Douspis \inst{6, \, 14} \and 
L.~Dumoulin \inst{15} \and 
X.~Dupac \inst{16} \and
P.~Filliatre \inst{17} \and 
P.~Fosalba \inst{11} \and
K.~Ganga \inst{18} \and 
F.~Gannaway \inst{2} \and 
B.~Gautier \inst{1} \and 
M.~Giard \inst{16} \and
Y.~Giraud--H{\'e}raud \inst{3, \, 24} \and 
R.~Gispert \inst{7\dag}\thanks{Richard Gispert passed away few weeks
after his return from the early mission to Trapani} \and 
L.~Guglielmi \inst{3, \, 24} \and
J.--Ch.~Hamilton \inst{3, \, 17} \and 
S.~Hanany \inst{19} \and
S.~Henrot--Versill{\'e} \inst{4} \and 
J.~Kaplan \inst{3, \, 24} \and
G.~Lagache \inst{7} \and 
J.--M.~Lamarre \inst{7,25} \and 
A.~E.~Lange \inst{8} \and 
J.~F.~Mac{\'\i}as--P{\'e}rez \inst{17} \and 
K.~Madet \inst{1} \and 
B.~Maffei \inst{2} \and
Ch.~Magneville \inst{5, \, 24} \and
D.~P.~Marrone \inst{19} \and
S.~Masi \inst{12} \and 
F.~Mayet \inst{5} \and 
A.~Murphy \inst{20} \and
F.~Naraghi \inst{17} \and 
F.~Nati \inst{12} \and
G.~Patanchon \inst{3, \, 24} \and
G.~Perrin \inst{17} \and 
M.~Piat \inst{7} \and 
N.~Ponthieu \inst{17} \and
S.~Prunet \inst{11} \and
J.--L.~Puget \inst{7} \and
C.~Renault \inst{17} \and 
C.~Rosset \inst{3, \, 24} \and
D.~Santos \inst{17} \and
A.~Starobinsky \inst{21} \and
I.~Strukov \inst{22} \and
R.~V.~Sudiwala \inst{2} \and 
R.~Teyssier \inst{11, \, 23} \and
M.~Tristram \inst{17} \and
C.~Tucker\inst{2} \and
J.--C.~Vanel \inst{3, \, 24} \and 
D.~Vibert \inst{11} \and 
E.~Wakui \inst{2} \and 
D.~Yvon \inst{5, \, 24}
}

   \offprints{reprints@archeops.org}
   \mail{benoit@archeops.org}
\institute{
%%1
Centre de Recherche sur les Tr{\`e}s Basses Temp{\'e}ratures,
BP166, 38042 Grenoble Cedex 9, France
\and
%%2
Cardiff University, Physics Department, PO Box 913, 5, The Parade,   
Cardiff, CF24 3YB, UK\and
%%3
Physique Corpusculaire et Cosmologie, Coll{\`e}ge de
France,  11 pl. M. Berthelot, F-75231 Paris Cedex 5, France
\and
%%4
Laboratoire de l'Acc{\'e}l{\'e}rateur Lin{\'e}aire, BP~34, Campus
Orsay, 91898 Orsay Cedex, France
\and
%%5
CEA-CE Saclay, DAPNIA, Service de Physique des Particules, 
Bat 141, F-91191 Gif sur Yvette Cedex, France
%%%%SPP/DAPNIA/DSM, CEA--CE Saclay, 91191 Gif-sur-Yvette  Cedex, France
\and
%%6
Laboratoire d'Astrophysique de l'Obs. Midi-Pyr{\'e}n{\'e}es,
14 Avenue E. Belin, 31400 Toulouse, France
\and
%%7
Institut d'Astrophysique Spatiale, B{\^a}t.  121, Universit{\'e} Paris
XI,
91405 Orsay Cedex, France
\and
%%8
California Institute of Technology, 105-24 Caltech, 1201 East 
California Blvd, Pasadena CA 91125, USA
\and
%%9
Jet Propulsion Laboratory, 4800 Oak Grove Drive, Pasadena, 
California 91109, USA
\and
%%10
IROE--CNR, Via Panciatichi, 64, 50127 Firenze, Italy
\and
%%11
Institut d'Astrophysique de Paris, 98bis, Boulevard Arago, 75014 Paris,
France
\and
%%12
Gruppo di Cosmologia Sperimentale, Dipart. di Fisica, Univ.  ``La
Sapienza'', P. A. Moro, 2, 00185 Roma, Italy
\and
%%13
Laboratoire d'Astrophysique, Obs. de Grenoble, BP 53, 
 38041 Grenoble Cedex 9, France
\and
%%14
Nuclear and Astrophysics Laboratory, Keble Road, Oxford,  OX1 3RH, UK
\and
%%15
CSNSM--IN2P3, B{\^a}t 108, 91405 Orsay Campus, France
\and
%%16
Centre d'{\'E}tude Spatiale des Rayonnements,
BP 4346, 31028 Toulouse Cedex 4, France
\and
%%17
Institut des Sciences Nucl{\'e}aires, 53 Avenue des Martyrs, 38026
Grenoble Cedex, France
\and
%%18
Infrared Processing and Analysis Center, Caltech, 770 South Wilson
Avenue, Pasadena, CA 91125, USA
\and
%%19
School of Physics and Astronomy, 116 Church St. S.E., University of
Minnesota, Minneapolis MN 55455, USA
\and
%%20
Experimental Physics, National University of Ireland, Maynooth, Ireland
\and
%%21
Landau Institute for Theoretical Physics, 119334 Moscow, Russia
\and
%%22
Space Research Institute, Profsoyuznaya St. 84/32, Moscow, Russia 
\and
%%23
CEA-CE Saclay, DAPNIA, Service d'Astrophysique, Bat 709, 
F-91191 Gif sur Yvette Cedex, France
\and
%%24
F{\'e}d{\'e}ration de Recherche APC, Universit{\'e} Paris 7, Paris, France
\and
%%25
LERMA, Observatoire de Paris, 61 Av. de l'Observatoire, 75014 Paris, France
}

%\date{\today} 
\date{Received 16 October 2002/Accepted 22 November 2002} 

\abstract{We analyze the cosmological constraints that
  Archeops (\cite{archeops_cl}) places on adiabatic cold dark matter
  models with passive power-law initial fluctuations.  Because its
  angular power spectrum has small bins in $\ell$ and large $\ell$
  coverage down to COBE scales, Archeops provides a precise
  determination of the first acoustic peak in terms of position at
  multipole $l_{\rm peak}=220\pm 6$, height and width.  An analysis of
  Archeops data in combination with other CMB datasets constrains the
  baryon content of the Universe, $\Ombhh =
  0.022^{+0.003}_{-0.004}$, compatible with Big-Bang nucleosynthesis
  and with a similar accuracy.  Using cosmological priors obtained
  from recent non--CMB data leads to yet tighter constraints on the
  total density, {\it e.g.}  $\OmT =1.00^{+0.03}_{-0.02}$
  using the HST determination of the Hubble constant.  An excellent
  absolute calibration consistency is found between Archeops and other
  CMB experiments, as well as with the previously quoted best fit
  model.  The spectral index $n$ is measured to be
  $1.04^{+0.10}_{-0.12}$ when the optical depth to reionization,
  $\tau$, is allowed to vary as a free parameter, and
  $0.96^{+0.03}_{-0.04}$ when $\tau$ is fixed to zero, both in good
  agreement with inflation.

\keywords{Cosmic microwave background --
    Cosmological parameters -- Early Universe -- Large--scale
    structure of the Universe} } 

\maketitle

%________________________________________________________________

\section{Introduction}

A determination of the amplitude of the fluctuations of the cosmic
microwave background (CMB) is one of the most promising techniques to
overcome a long standing problem in cosmology -- setting constraints
on the values of the cosmological parameters. Early detection of a
peak in the region of the so-called first acoustic peak ($\ell \approx
200$) by the Saskatoon experiment (\cite{netterfield97}), as well as
the availability of fast codes to compute theoretical amplitudes
(\cite{seljak}) has provided a first constraint on the geometry of the
Universe (\cite{lineweaver}, \cite{hancock}). The spectacular results
of Boomerang and Maxima have firmly established the fact that the
geometry of the Universe is very close to flat (\cite{boom},
\cite{maxima}, \cite{lange}, \cite{balbi}). Tight constraints on most
cosmological parameters are anticipated from the Map (\cite{map}) and
Planck (\cite{planck}) satellite experiments. Although experiments
have already provided accurate measurements over a wide range of
$\ell$, degeneracies prevent a precise determination of some
parameters using CMB data alone. For example, the matter content
$\Omm$ cannot be obtained independently of the Hubble constant.
Therefore, combinations with other cosmological measurements (such as
supernov{\ae}, Hubble constant, and light element fractions) are used
to break these degeneracies. Multiple constraints can be obtained on
any given parameter by combining CMB data with anyone of these other
measurements. It is also of interest to check the consistency between
these multiple constraints.  In this letter, we derive constraints on
a number of cosmological parameters using the measurement of CMB
anisotropy by the Archeops experiment (\cite{archeops_cl}).  This
measurement provides the most accurate determination presently
available of the angular power spectrum at angular scales of the first
acoustic peak and larger.

\begin{figure}[t]
\resizebox{\hsize}{!}{\includegraphics[clip]{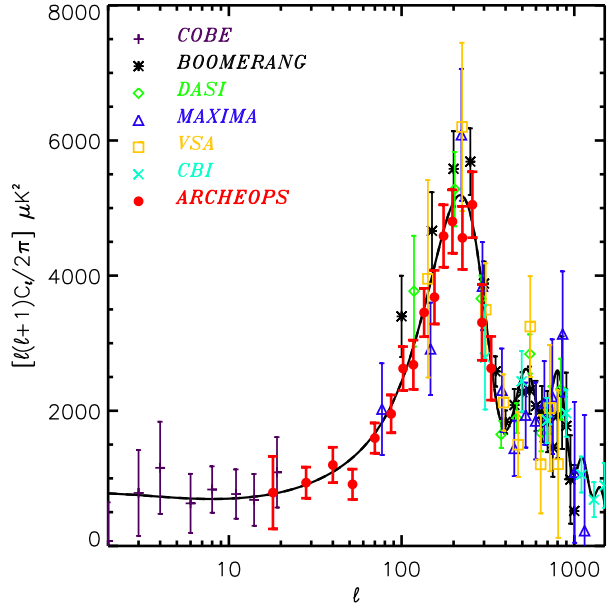}}
\caption{\label{powplot1}Measurements of the CMB angular power
spectrum by Archeops (in red dots) compared with CBDMVC datasets. 
A $\Lambda$CDM model (see text for parameters) is overplotted and
appears to be in good agreement with all the data.}
\end{figure}

%________________________________________________________________
\section{Archeops angular power spectrum}

\begin{figure}[!ht]
\begin{center}
\resizebox{\hsize}{!}{\includegraphics{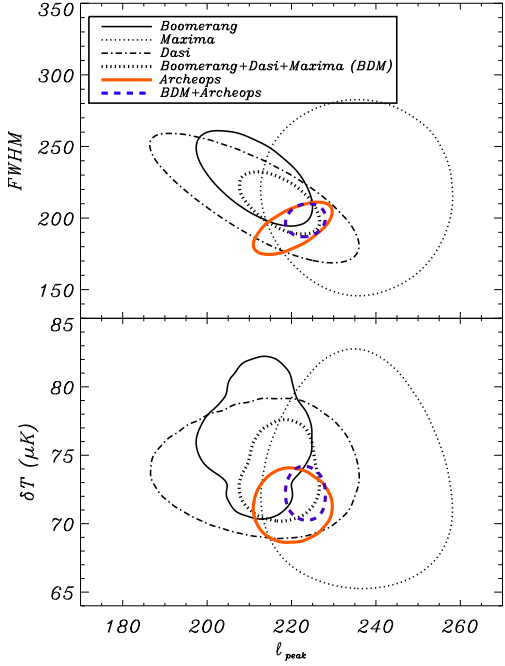}}\end{center}
\caption{\label{compeak} Gaussian fitting of the first acoustic
  peak using Archeops and other CMB experiments ($\ell\le 390$). {\sl
    Top~panel}: 68\%~CL likelihood contours in the first peak position
  and FWHM ($\ell_{\rm peak}, FWHM$) plane; {\sl Bottom~panel}: 68\%~CL
  likelihood contours in the first peak position and height
  ($\ell_{\rm peak}, \delta T_{\rm peak}$) plane for different CMB experiments
  and combinations. The width of the peak is constrained differently
  by Archeops and BDM experiments, so that the intersection lies on
  relatively large $\ell_{\rm peak}$.  Hence, the BDM~+~Archeops zone is
  skewed to the right in the bottom panel.}
\end{figure}

The first results of the February~2002 flight of Archeops are detailed
in \cite{archeops_cl}. The band powers used in this analysis are
plotted in Fig.~\ref{powplot1} together with those of other
experiments (CBDMVC for COBE, Boomerang, Dasi, Maxima, VSA, and CBI;
\cite{tegmark}, \cite{boom2}, \cite{halverson}, \cite{maxima2}, 
\cite{scott}, \cite{pearson}).  Also plotted is a $\Lambda$CDM model
(computed using CAMB, \cite{lewis}), with the following cosmological
parameters: $\Theta=(\OmT, \OmL, \Ombhh, h, n, Q,
\tau)=(1.00,0.7,0.02,0.70,1.00,18\mu K,0.)$ where the parameters are
the total energy density, the energy density of a cosmological
constant, the baryon density, the normalized Hubble constant
($H_0=100\, h \rm\, km/s/Mpc$), the spectral index of the scalar
primordial fluctuations, the normalization of the power spectrum and
the optical depth to reionization, respectively.  The predictions of
inflationary motivated adiabatic fluctuations, a plateau in the power
spectrum at large angular scales followed by a first acoustic peak,
are in agreement with the results from Archeops and from the other
experiments.  Moreover, the data from Archeops alone provides a
detailed description of the power spectrum around the first peak.  The
parameters of the peak can be studied without a cosmological prejudice
(\cite{knox,dousfer}) by fitting a constant term, here fixed to match
COBE amplitude, and a Gaussian function of $\ell$.  Following this
procedure and using the Archeops and COBE data only, we find
(Fig.~\ref{compeak}) for the location of the peak $\ell_{\rm peak}=220\pm
6$, for its width $FWHM =192\pm 12$, and for its amplitude $\delta
T=71.5\pm 2.0\,\rm \mu K$ (error bars are smaller than the calibration
uncertainty from Archeops only, because COBE amplitude is used for the
constant term in the fit).  This is the best determination of the
parameters of the first peak to date, yet still compatible with other
CMB experiments.

\renewcommand{\arraystretch}{1.4}
\begin{table}[!hb]
\begin{center}
\begin{tabular}{cccccccc}
\hline
\hline
 & $\OmT$&$\OmL$&$\Ombhh $&$h$&$n$&$Q$&$\tau$\\
\hline
Min. & 0.7    & 0.0  & 0.00915  & 0.25   & 0.650 &  11 &  0.0 \\
Max. & 1.40   & 1.0  & 0.0347   & 1.01   & 1.445 &  27 &  1.0 \\
Step & 0.05   & 0.1  & 0.00366  & *1.15  & 0.015 & 0.2 &  0.1 \\
\hline
\end{tabular}
\end{center}  
\caption{\label{grid_table}The grid of points in the 7~dimensional 
space of cosmological models that was used to set constraint on 
the cosmological parameters. $^*$For $h$ we adopt a logarithmic binning:
$h(i+1)=1.15\cdot h(i)$ ; $Q$ is in $\mu K$.}
\end{table}

%---------------------------------------------------------------------------
\section{Model grid and likelihood method}

To constrain cosmological models we constructed a $4.5\times 10^8$
$C_\ell$ database.  Only inflationary motivated models with adiabatic
fluctuations are being used. The ratio of tensor to scalar modes is
also set to zero.  As the hot dark matter component modifies mostly
large $\ell$ values of the power spectrum, this effect is neglected in
the following.  Table~\ref{grid_table} describes the corresponding
gridding used for the database. The models including reionization have
been computed with an analytical approximation (\cite{griffiths}).

Cosmological parameter estimation relies upon the knowledge of the
likelihood function $\mathcal{L}$ of each band power estimate.
Current Monte Carlo methods for the extraction of the $C_\ell$
naturally provide the distribution function $\mathcal{D}$ of these
power estimates.  The analytical approach described in \cite{douspis2}
and \cite{BDBL} allows to construct the needed $\mathcal{L}$ in an
analytical form from $\mathcal{D}$. Using such an approach was proven
to be equivalent to performing a full likelihood analysis on the maps.
Furthermore, this leads to unbiased estimates of the cosmological
parameters (\cite{wandelt}, \cite{bond}, \cite{douspis1a}), unlike other
commonly used $\chi^2$ methods. In these methods, $\mathcal{L}$ is
also assumed to be Gaussian. However this hypothesis is not valid,
especially for the smaller modes covered by Archeops. The difference
between our well--motivated shape and the Gaussian approximation
induces a 10\% error in width for large--scale bins.  The parameters
of the analytical form of the band power likelihoods $\mathcal{L}$
have been computed from the distribution functions of the band powers
listed in Tab.~1 of \cite{archeops_cl}. Using $\mathcal{L}$, we
calculate the likelihood of any of the cosmological models in the
database and maximize the likelihood over the 7\% calibration
uncertainty. We include the calibration uncertainty of each experiment
as extra parameters in our analysis.  The prior on these parameters
are taken as Gaussians centered on unity, with a standard deviation
corresponding to the quoted calibration uncertainty of each dataset.
The effect of Archeops beam width uncertainty, which leads to less
than 5\% uncertainty on the $C_\ell$'s at $\ell \le 350$, is
neglected.

\begin{figure}[!b]
\resizebox{\hsize}{!}{\includegraphics[clip]{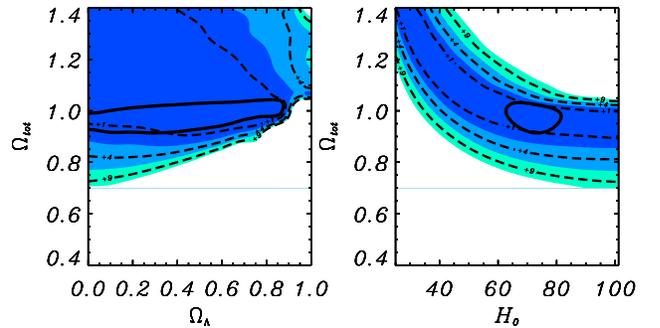}}
\caption{\label{arcmix}Likelihood contours in the $(\OmL, \OmT)$
  (left) and $(H_0, \OmT)$ (right) planes using the Archeops dataset;
  the three colored regions (three contour lines) correspond to resp.
  68, 95 and 99\% confidence levels for 2-parameters (1-parameter)
  estimates. Black solid line is given by the combination
  Archeops + HST, see text.}

\end{figure}    

A numerical compilation of all the results is given in
Tab.~\ref{result_table}.  Some of the results are also presented as
2-D contour plots, showing in shades of blue the regions where the
likelihood function for a combination of any two parameters drops to
68\%, 95\%, and 99\% of its initial value. These levels are computed
from the minimum of the negative of the log likelihood plus $\Delta
=2.3,\; 6.17\; \rm and\; 11.8$. They would correspond to 1, 2, 3
$\sigma$ respectively if the likelihood function was Gaussian.  Black
contours mark the limits to be projected if confidence intervals are
sought for any one of the parameters.  To calculate either 1-- or 2--D
confidence intervals, the likelihood function is maximized over the
remaining parameters.  All single parameter confidence intervals that
are quoted in the text are $1\,\sigma$ unless otherwise stated, and we
use the notation $\chi^2_{gen}= m/n$ to mean that the generalized
$\chi^2$ has a value $m$ with $n$ degrees of freedom. In all cases
described below we find models that do fit the data and therefore
confidence levels have a well defined statistical interpretation.
\cite{douspis2} describes how to evaluate the goodness of fit and
Tab.~\ref{result_table} gives the various $\chi^2$ values. When we use
external non--CMB priors on some of the cosmological parameters, the
analysis is done by multiplying our CMB likelihood hypercube by
Gaussian shaped priors with mean and width according to the published
values.

%---------------------------------------------------------------------------
\section{Cosmological Parameter constraints}

\subsection{Archeops}\label{archonly}

We first find constraints on the cosmological parameters using the
Archeops data alone. The cosmological model that presents the best fit
to the data has a $\chi^2_{gen}= 6/9$. Figure~\ref{arcmix} gives
confidence intervals on different pairs of parameters.  The Archeops
data constrain the total mass and energy density of the Universe
($\OmT$) to be greater than 0.90, but it does not provide
strong limits on closed Universe models.  Fig.~\ref{arcmix} also shows
that $\OmT$ and $h$ are highly correlated (\cite{douspis1b}).
Adding the HST constraint for the Hubble constant, $H_0=72\pm8
\rm\,km/s/Mpc$ (68\% CL, \cite{freedman}), leads to the tight
constraint $\OmT = 0.96^{+0.09}_{-0.04}$ (full line in
Fig.~\ref{arcmix}), indicating that the Universe is flat.

Using Archeops data alone we can set significant constraints neither on
the spectral index $n$ nor on the baryon content $\Ombhh$
because of lack of information on fluctuations at small angular
scales.

\subsection{COBE, Archeops, CBI}

\begin{figure}[!t]
\resizebox{\hsize}{!}{\includegraphics[clip]{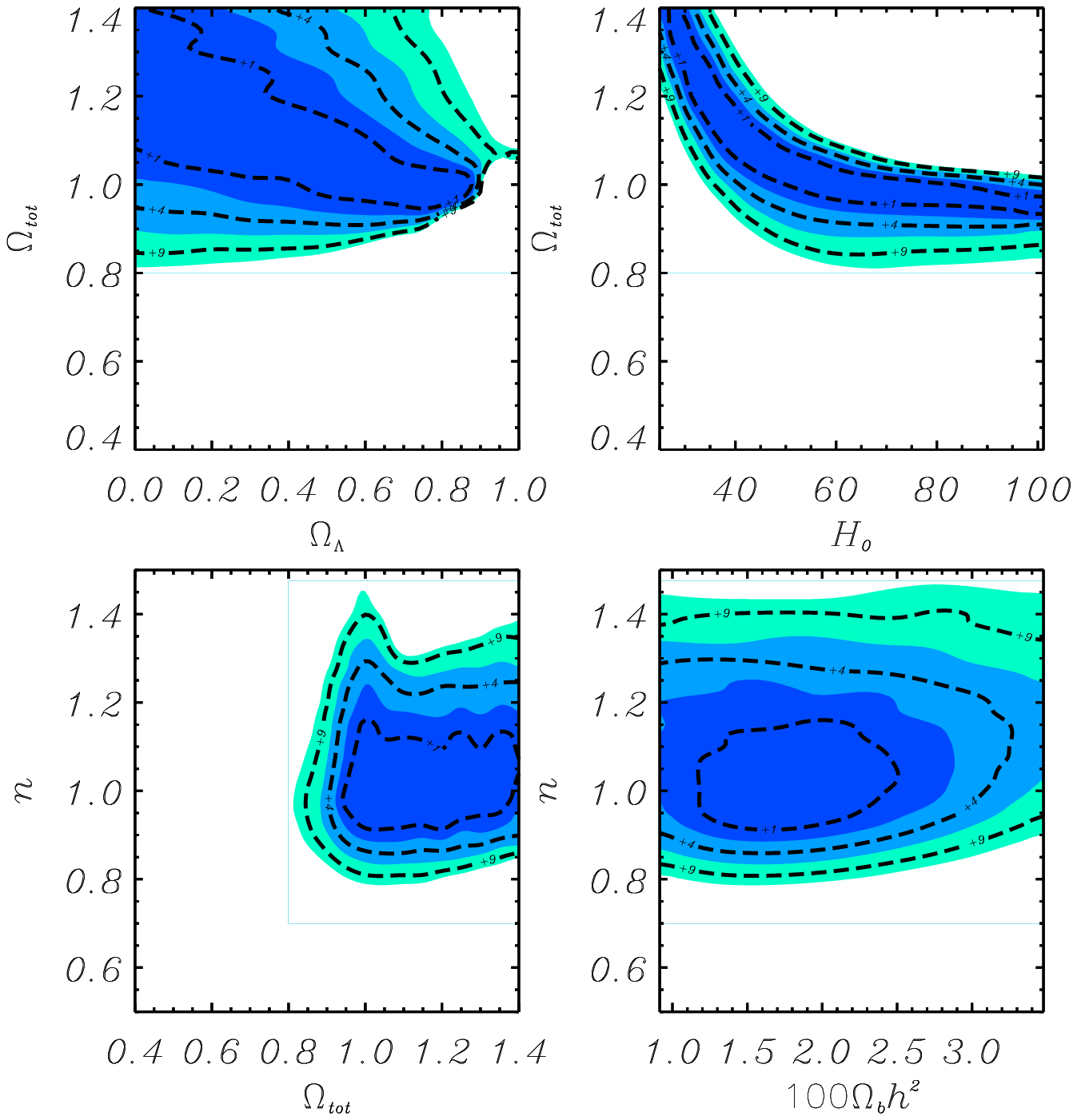}}
\caption{\label{ccamix}Likelihood contours for 
  (COBE + Archeops + CBI) in the $(\OmL, \OmT)$,
  $(H_0, \OmT)$, $(\OmT, n)$ and ($\Ombhh,n$) planes.}
\end{figure}

We first combine only COBE/DMR, CBI and Archeops so as to include
information over a broad range of angular scales, $2 \le \ell \le
1500$, with a minimal number of experiments\footnote{For CBI data, we
  used only the joint mosaic band powers and restrict ourselves to
  $\ell\le 1500$.}.  The results are shown in Fig.~\ref{ccamix}, with
a best model $\chi^2_{gen}= 9/20$.  The constraint on open models is
stronger than previously, with a total density $\OmT = 1.16
^{+0.24}_{-0.20}$ at 68\% CL and $\OmT > 0.90$ at 95\% CL. The
inclusion of information about small scale fluctuations provides a
constraint on the baryon content, $\Ombhh = 0.019^{+0.006}_{-0.007}$ in
good agreement with the results from BBN (\cite{omeara}: $\Ombhh =
0.0205\pm0.0018$). The spectral index
$n=1.06^{+0.11}_{-0.14}$ is compatible with a scale invariant
Harrison--Zel'dovich power spectrum.

\subsection{Archeops and other CMB experiments}    

\begin{figure}[!ht]
\resizebox{\hsize}{!}{\includegraphics[clip]{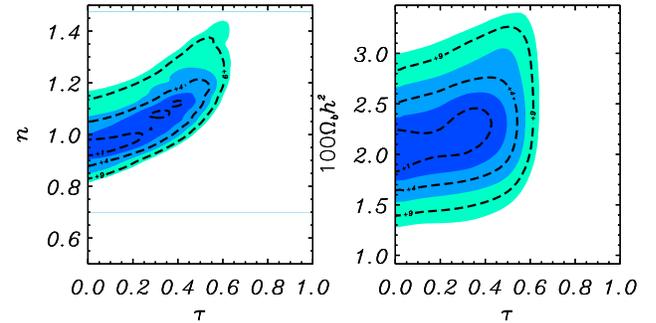}}
\caption{\label{tau}Likelihood contours in the $(\tau, n)$ and $(\tau,
\Ombhh)$ planes using Archeops + CBDMVC datasets.}
\end{figure}

By adding the experiments listed in Fig.~\ref{powplot1} we now provide
the best current estimate of the cosmological parameters using CMB
data only.  The constraints are shown on Fig.~\ref{tau} and
\ref{alllow}~(left).  The combination of all CMB experiments provides
$\sim 10\%$ errors on the total density, the spectral index and the
baryon content respectively: $\OmT=1.15^{+0.12}_{-0.17}$,
$n=1.04^{+0.10}_{-0.12}$ and $\Ombhh=0.022^{+0.003}_{-0.004}$.  These
results are in good agreement with recent analyses performed by other
teams (\cite{boom2}, \cite{pryke}, \cite{rubino}, \cite{sievers},
\cite{wang}). One can also note that the parameters of the
$\Lambda$CDM model shown in Fig.~\ref{powplot1} are included in the
68\%~CL contours of Fig.~\ref{alllow}~(right).

As shown in Fig.~\ref{tau} the spectral index and the optical depth
are degenerate. Fixing the latter to its best fit value, $\tau = 0$,
leads to stronger constraints on both $n$ and $\Ombhh$.  With this
constraint, the prefered value of $n$ becomes slightly lower than 1,
$n=0.96^{+0.03}_{-0.04}$, and the constraint on $\Ombhh$ from CMB alone
is not only in perfect agreement with BBN determination but also has
similar error bars, $\Ombhh_{\rm (CMB)}=0.021^{+0.002}_{-0.003}$.  It is
important to note that many inflationary models (and most of the
simplest of them) predict a value for $n$ that is slightly less than
unity (see, e.g.,~\cite{Linde} and \cite{lyth} for a recent review).

\subsection{Adding non--CMB priors}

\begin{figure}[!ht]
\resizebox{\hsize}{!}{\includegraphics[clip]{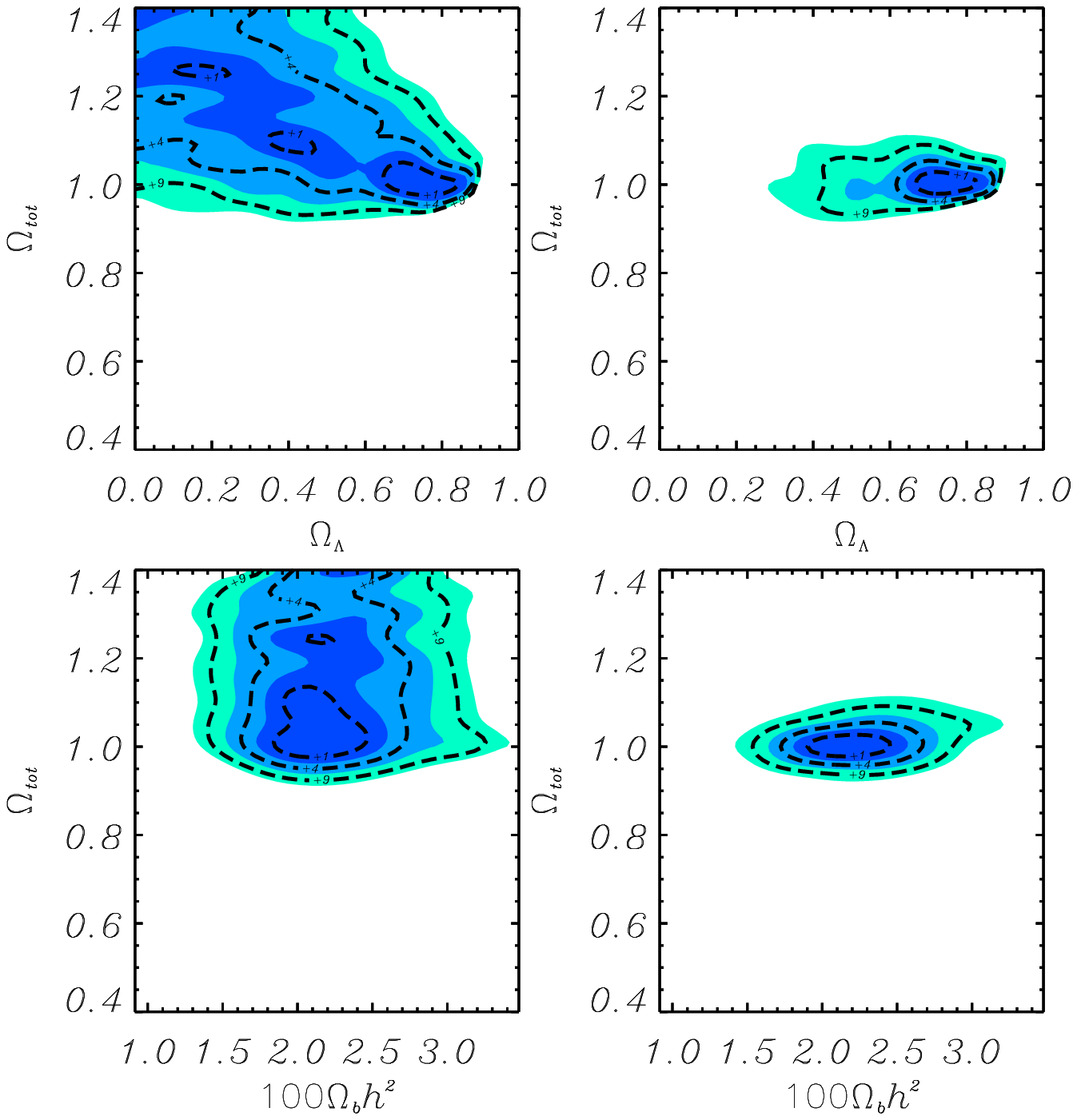}}
\caption{\label{alllow}Likelihood contours in the $(\OmT, \OmL)$ and
 $(\OmT, \Ombhh)$ planes. Left: constraints using Archeops+CBDMVC
 datasets. Right: adding HST prior for $H_0$.}
\end{figure}    

\begin{figure}[!ht]
\resizebox{\hsize}{!}{\includegraphics[clip]{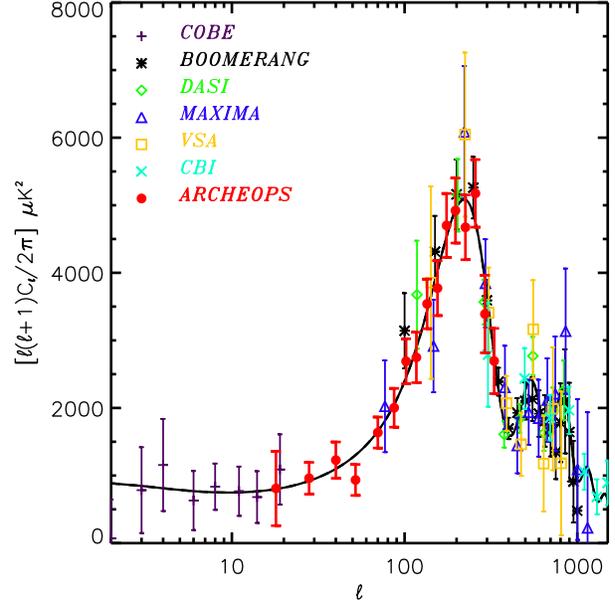}}
\caption{\label{powplot2} Best model obtained from the Archeops~+~CBDMVC~+~HST
  analysis with recalibrated actual datasets. The fitting allowed the
  gain of each experiment to vary within their quoted absolute
  uncertainties.  Recalibration factors, in temperature, which are applied in
  this figure, are 1.00, 0.96, 0.99, 1.00, 0.99, 1.00, and 1.01, for
  COBE, Boomerang, Dasi, Maxima, VSA, CBI and Archeops respectively,
  well within 1~$\sigma$ of the quoted absolute uncertainties ($<1$,
  10, 4, 4, 3.5, 5 and 7\%).}
\end{figure}

In order to break some degeneracies in the determination of
cosmological parameters with CMB data alone, priors coming from other
cosmological observations are now added.  First we consider priors
based on stellar candles like HST determination of the Hubble constant
(\cite{freedman}) and supernov{\ae} determination of $\Omm$ and
$\Lambda$ (\cite{perlmutter}). We also consider non stellar
cosmological priors like BBN determination of the baryon content,
(\cite{omeara}), and baryon fraction determination from X-ray clusters
(\cite{roussel}, \cite{sadat}).  For the baryon fraction we use a low
value, BF(L), $f_b = 0.031h^{-3/2} +0.012\; (\pm 10\%)$, and a high
value, BF(H), $f_b = 0.048h^{-3/2} + 0.014\; (\pm 10\%)$
(\cite{douspis1b} and references therein).  The results with the HST
prior are shown in Figure~\ref{alllow}~(right). Considering the
particular combination Archeops~+~CBDMVC~+~HST, the best fit model,
within the Tab.~\ref{grid_table} gridding, is $(\OmT,
\OmL, \Ombhh, h, n, Q,
\tau)=(1.00,0.7,0.02,0.665,0.945,19.2{\rm \mu K},0.)$ with a
$\chi^2_{gen}=41/68$. The model is shown in Fig.~\ref{powplot2} with
the data scaled by their best--fit calibration factors which were
simultaneously computed in the likelihood fitting process. The
constraints on $h$ break the degeneracy between the total matter
content of the Universe and the amount of dark energy as discussed in
Sect.~\ref{archonly}.  The constraints are then tighter as shown in
Fig.~\ref{alllow}~(right), leading to a value of $\OmL =
0.73^{+0.09}_{-0.07}$ for the dark energy content, in agreement with
supernov{\ae} measurements if a flat Universe is assumed.
Table~\ref{result_table} also shows that Archeops~+~CBDMVC
cosmological parameter determinations assuming either $\OmT=1$ or the
HST prior on $h$ are equivalent at the 68\%~CL.

\renewcommand{\arraystretch}{1.2}
\begin{table*}
\bc
\begin{tabular}{c|ccccccc}
\hline
\hline
Data & $\OmT$ & $n_s$ & $\Ombhh$ & $h$ &
$\OmL$ & $\tau$ & $\chi^2_{gen}/dof$\\
\hline
Archeops 
& $ > 0.90$ 
& $1.15^{+0.30}_{-0.40}$ 
& $ --  $  
& $ -- $ 
& $ <0.9$  
& $ <0.45$  
& $  6/9 $\\
\hline
Archeops + COBE + CBI 
& $1.16^{+0.24}_{-0.20}$ 
& $1.06^{+0.11}_{-0.14}$ 
& $0.019^{+0.006}_{-0.007}$ 
& $ >0.25 $  
& $ < 0.85 $ 
& $ <0.45$  
& $  9/20 $\\
\hline
CMB 
& $1.18^{+0.22}_{-0.20}$ 
& $1.06^{+0.14}_{-0.20}$ 
& $0.024^{+0.003}_{-0.005}$ 
& $0.51^{+0.30}_{-0.30}$ 
& $< 0.85$  
& $ <0.55$  
& $ 37/52$ \\
Archeops + CMB 
& $1.15^{+0.12}_{-0.17}$ 
& $1.04^{+0.10}_{-0.12}$ 
& $0.022^{+0.003}_{-0.004}$ 
& $0.53^{+0.25}_{-0.13}$ 
& $< 0.85$  
& $ <0.4$  
& $ 41/67$ \\
Archeops + CMB + $\tau=0$
& $1.13^{+0.12}_{-0.15}$ 
& $0.96^{+0.03}_{-0.04}$ 
& $0.021^{+0.002}_{-0.003}$ 
& $0.52^{+0.20}_{-0.12}$ 
& $< 0.80$  
& $ 0.0$  
& $ 41/68$ \\
Archeops + CMB + $\OmT=1$
& $1.00$
& $1.04^{+0.10}_{-0.12}$ 
& $0.021^{+0.004}_{-0.003}$ 
& $0.70^{+0.08}_{-0.08}$ 
& $0.70^{+0.10}_{-0.10}$ 
& $ < 0.40$  
& $ 41/68$ \\
\hline
Archeops + CMB + HST
& $1.00^{+0.03}_{-0.02}$ 
& $1.04^{+0.10}_{-0.08}$ 
& $0.022^{+0.003}_{-0.002}$ 
& $0.69^{+0.08}_{-0.06}$ 
& $0.73^{+0.09}_{-0.07}$  
& $ <0.42$  
& $ 41/68$ \\
Archeops + CMB + HST + $\tau=0$
& $1.00^{+0.03}_{-0.02}$ 
& $0.96^{+0.02}_{-0.04}$ 
& $0.021^{+0.001}_{-0.003}$ 
& $0.69^{+0.06}_{-0.06}$ 
& $0.72^{+0.08}_{-0.06}$  
& $ 0.0$  
& $ 41/69$ \\
Archeops + CMB + SN1a
& $1.04^{+0.02}_{-0.04}$ 
& $1.04^{+0.10}_{-0.12}$ 
& $0.022^{+0.003}_{-0.004}$ 
& $0.60^{+0.10}_{-0.07}$ 
& $0.67^{+0.11}_{-0.03}$  
& $ <0.40$  
& $ 41/69$ \\
Archeops + CMB + BBN
& $1.12^{+0.13}_{-0.14}$ 
& $1.04^{+0.10}_{-0.12}$ 
& $0.020^{+0.002}_{-0.002}$ 
& $0.50^{+0.15}_{-0.10}$ 
& $< 0.80$  
& $ <0.25$  
& $ 41/68$ \\
Archeops + CMB + BF(H)
& $1.11^{+0.12}_{-0.11}$ 
& $1.03^{+0.12}_{-0.14}$ 
& $0.022^{+0.004}_{-0.004}$ 
& $0.46^{+0.09}_{-0.11}$ 
& $0.45^{+0.10}_{-0.10} $  
& $ <0.40$  
& $ 43/69$ \\
Archeops + CMB + BF(L)
& $1.22^{+0.18}_{-0.12}$ 
& $1.03^{+0.07}_{-0.13}$ 
& $0.021^{+0.003}_{-0.004}$ 
& $< 0.40 $ 
& $ < 0.3 $  
& $ <0.40$  
& $ 45/69$ \\
\hline
\end{tabular}
\caption{\label{result_table}
Cosmological parameter constraints from
combined datasets. Upper and lower limits are given for 68\% CL.
See text for details on priors. The
central values are given by the mean of the likelihood. 
The quoted error bars are at times smaller than the
parameter grid spacing, and are thus in fact determined by an
interpolation of the likelihood function between adjacent grid points.
}
\ec
\end{table*}

\section{Conclusion}

Constraints on various cosmological parameters have been derived by
using the Archeops data alone and in combination with other
measurements.  The measured power at low $\ell$ is in agreement with
the COBE data, providing for the first time a direct link between the
Sachs--Wolfe plateau and the first acoustic peak.  The Archeops data
give a high signal-to-noise ratio determination of the parameters of
the first acoustic peak and of the power spectrum down to COBE scales
($\ell=15$), because of the large sky coverage that greatly reduces
the sample variance.  The measured spectrum is in good agreement with
that predicted by simple inflation models of scale--free adiabatic
peturbations.  Archeops on its own also sets a constraint on open
models, $\OmT > 0.90$ (68\%~CL). In combination with CBDMVC
experiments, tight constraints are shown on cosmological parameters
like the total density, the spectral index and the baryon content,
with values of $\OmT = 1.13^{+0.12}_{-0.15}$, $n =
0.96^{+0.03}_{-0.04}$ and $\Ombhh = 0.021^{+0.002}_{-0.003}$
respectively, all at 68\%~CL and assuming $\tau=0$. These results lend
support to the inflationary paradigm. The addition of non--CMB
constraints removes degeneracies between different parameters and
allows to achieve a $10\%$ precision on $\Ombhh$ and $\OmL$
and better than 5\% precision on $\OmT$ and $n$. Flatness of the
Universe is confirmed with a high degree of precision: $\OmT =
1.00^{+0.03}_{-0.02}$ (Archeops~+~CMB~+~HST).

%________________________________________________________________

\begin{acknowledgements}
  The authors would like to thank the following institutes for funding
  and balloon launching capabilities: CNES (French space agency), PNC
  (French Cosmology Program), ASI (Italian Space Agency), PPARC, NASA,
  the University of Minnesota, the American Astronomical Society and a
  CMBNet Research Fellowship from the European Commission.
\end{acknowledgements}

%________________________________________________________________

\end{document}